# ANALYSE OF WOVEN REINFORCEMENT PREFORMING BY EXPERIMENTAL APPROACH

**D. Soulat [1], S. Allaoui [1], S. Chatel [2]**

[1] Institut PRISME/MMH, Université d'Orléans, France, damien.soulat@univ-orleans.fr

[2] EADS France, Innovation Works, Suresnes , France

*Abstract:* An experimental device of dry preform stamping was designed and carried out for the first stage of the RTM process. This tool was developed to test the feasibility to obtain specific double curved shape constituted with dry fabric reinforcement. Optical strain measurement can quantify defects on the composite piece. It is just enough to change the desired punch-die set of the preform. This tool could be used to validate numerical simulation of the process.

## 1. INTRODUCTION

The use of fiber fabrics in transportation industries is increasing because it gives the possibility to reach complex shapes, for a lighter final product with the R.T.M. process. This process consists of a drawing operation of the fabric before a resin is injected. [1]. Aeronautical industries interrogation is on the feasibility to obtain with this process one 3D very complex double-curved shape (figure1) for thick piece constituted by several ply of fabric and without defects.

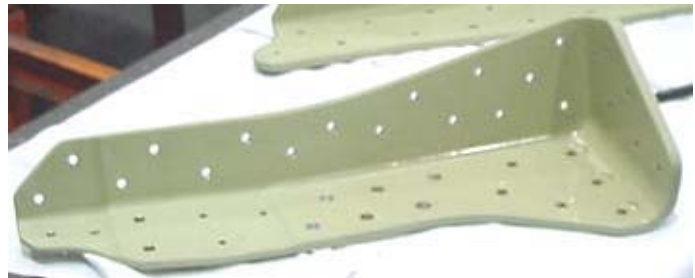

**Figure 1:** Aeronautical application

The aim of this paper is on the development of an experimental tool for the first stage of this RTM process, in collaboration with the EADS Company. This tool must permit to envisage the feasibility of a process without defect for thick fabric but also to know the directions of the reinforcements after shaping. These directions condition strongly the mechanical behaviour of the final textile composite structure. In addition, the angles between warp and weft yarns influence the permeability of the reinforcement and thus the filling of the resin in the case of a liquid moulding process [2, 3]. For another point of view this tool will permit to validate numerical simulation of this shaping stage realized by LAMCOS Laboratory. These numerical simulations [4, 5], by mechanical analysis of the fabric deformation takes into account the specific mechanical behaviour of the fabric used and especially loads (for example on the blank-holder) which mainly influence the result and the quality of the piece obtained. Consequently this experimental tool is instrumented by systems for strain measurement in yarns, and we choose by optical method, and second to measure parameters process, like pressure or punch velocity.

## 2. PRESENTATION OF THE TOOL

An experimental tool was developed in order to follow the behaviour of dry fabric during the first stage of RTM process.

### 2.1 Process parameters instrumentation

This forming testing machine is composed of a electric cylinder. Different sets of die-punch can be assembled on this tool. Figure 2 shows the different punch associated of these shapes. These shapes were considered one because they permit to obtain double-curved shape. On the other hand, on each face, edge, or angular point of these shapes, in function of the initial orientation of yarn, the fabric deformability, and consequently the defaults of forming could be different  A blank holder system, control by several pneumatic cylinder is present to tighten at different places of the fabric tested. Some thickness, or plies number of fabric could be tested. In addition various devices were added on this tool. One of them makes it possible to apply an initial tension along yarns of the fabric before this forming step. This device is necessary for very thick fabrics or for several plies considered. Naturally it's possible to measure all parameters which could affect on the process and consequently on the shape obtained, like stroke or load of the punch, blank holder pressure, orientation of the punch, initial tension..

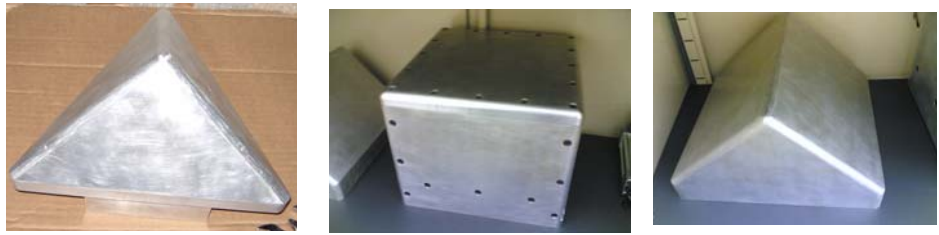

**Figure 2:** Punches geometry.

### 2.2 Optical strain measurements

Optical strain measurements are made simultaneously with two numerical cameras at macro-scale (pictures of the whole shape) and at mesoscale on some few woven cells or at micro-scale within a yarn. They permit measurement independent of the tool in fact to understand the internal behaviour of the fabric during shaping. The mechanical behaviour which results is complex and specific considering the possibilities of relative movements between yarns. The strain fields are computed using which used marks tracking technique [6, 7] from the displacements of points marked on fabrics. The macroscopic measures permit to compute the shape obtained for the each punch used and during all the process, which could be compared with the shape computed by numerical simulation. Position of the yarn during the process is another important parameter for the simulation in fact to apply the mechanical law behaviour in tension along their direction (warp and weft). During shaping without resin, the principal deformation mode of woven fabrics [8] is the plane shear and could be estimated by measurement of angle between yarns, during the process. When this angle is too large, shear stiffness become significant and it can involve formation of wrinkles [9]. All these quantities permit to quantify defects during this first stage of the RTM process.

## 3. EXPERIMENTAL RESULTS

The tests presented in this paper are carried out on a composite woven reinforcement used in aeronautics. It is denoted G1151® fabric and constituted (figure 3) by an interlock weaving links two carbon plies (630g/m²).

### 3.1 Global preform analysis.

When the forming test is finished, the fabric preform is removed from the tools and fixed with an epoxy resin spray. On figure 4 a, we present for square punch used, the preform obtained for one ply of the carbon reinforcement considered. In good adequation with the punch shape, the preform is symmetrical. We can observe that the reinforcement is well-stretched; on faces, edges or corner of this shape no defaults like tears appear. We can conclude that blank holder pressure is well-chosen in function of the punch's stroke.

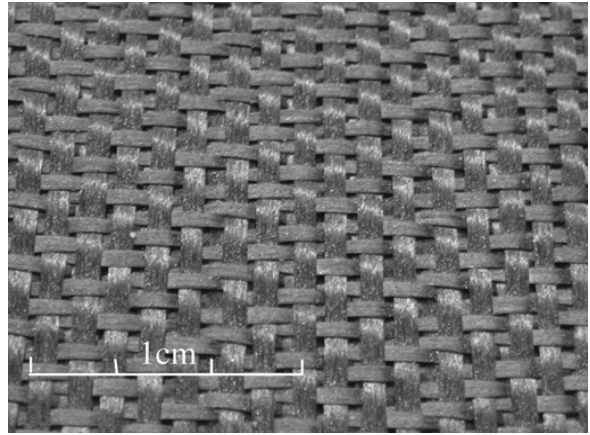

**Figure 3:** Woven carbon reinforcement (G1151®)

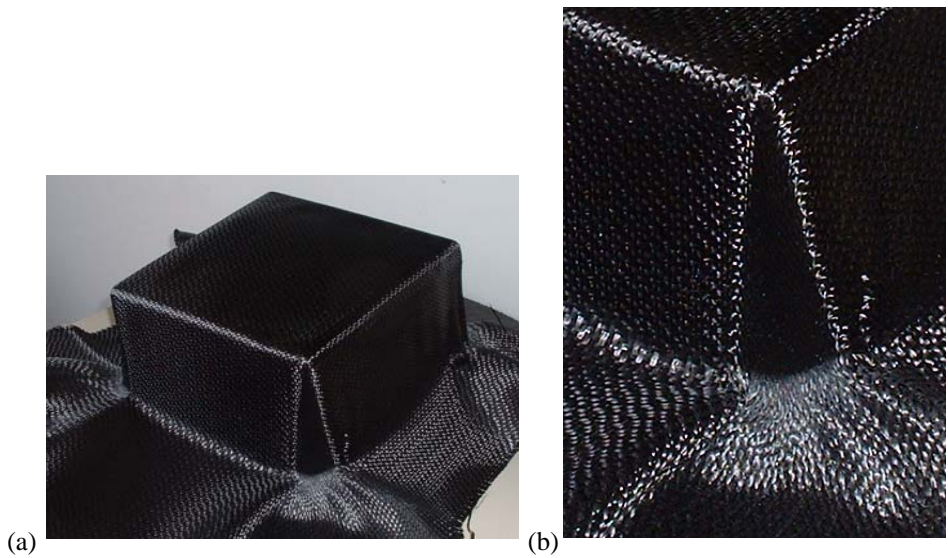

(a)                                                                                      (b)

**Figure 4:** Global preform obtained. (b) Local shear strain on edge.

The deformability of the fabric is possible by rotation between yarns, and we can notice (figure 4.b), for this shape, locally (corner, edges) that this rotation could become very important. Consequently on these edges, for example, it will be necessary to compute exactly the angle in fact to estimate the shear stiffness in adequacy with the shear behaviour of this fabric [9].

**3.2 Optical measurement analyses**.

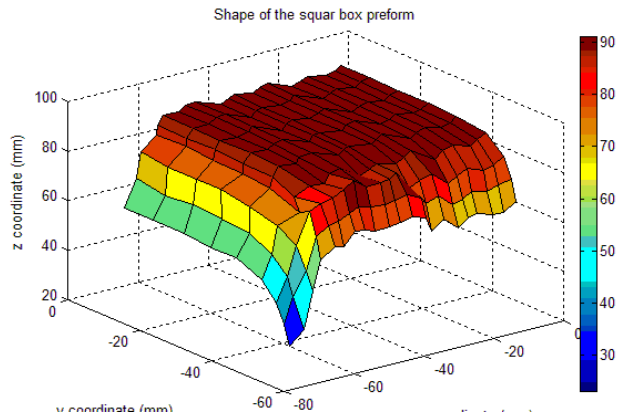

**Figure 5:** Shape of the square box preform after stamping

**3**

With the software of image processing it makes it possible to obtain the instantaneous position of each point marked on fabric. These positions are obtained with the three space co-ordinates. Thus, the post processing enables us to reconstruct the 3D geometry of the preform (Fig. 5). The software also gives displacements, and could compute plane component of strain for the fabric. Knowing the instantaneous positions, the shear angles between yarns can thus be computed. Indeed, the local behaviour of fibres can be studied for the same test by considering only the markers along on these fibres. Figure 6 shows an example of the evolution of position of three fibres between the initial and final position. The area considered, is at the top of the face. We noted that the evolution of the position of a fibre is more important near to the corner of the shape.

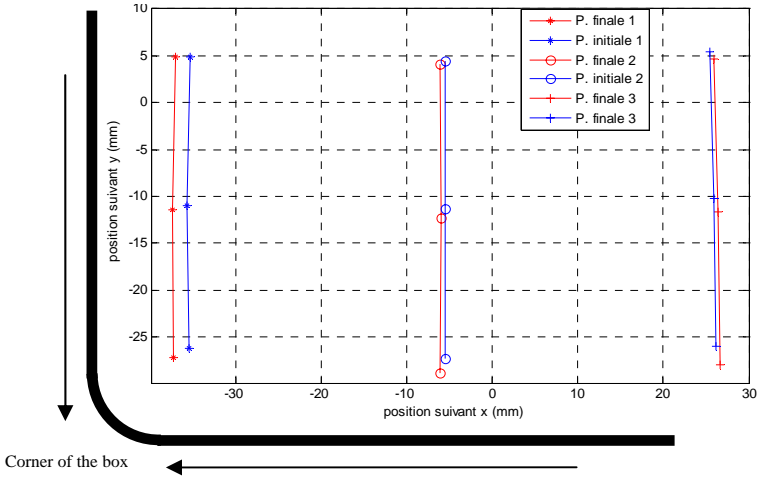

**Figure 6:** Evolution of the fibres positions

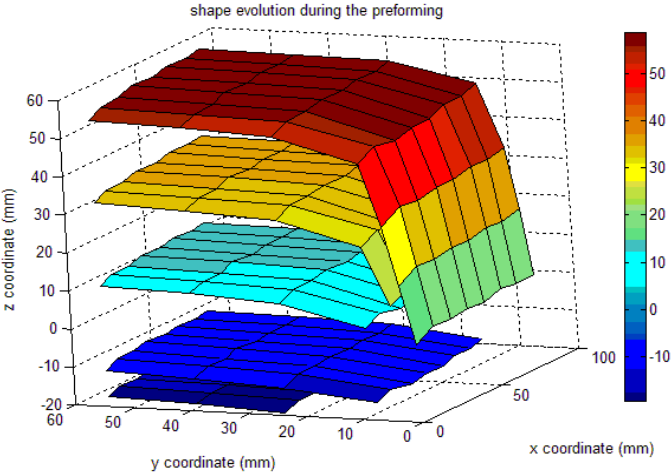

**Figure 7:** Shape evolution of an area of a square box preform during stamping

However, the data acquisition is done during all the process, enables us to analyse the behaviour of the fabric at any moment of the test, for each shape considered. Thus, figure 7 represents the evolution of the geometry shape at different stage of the stamping. This possibility can be used to determine when defaults appears in the reinforcement or for which process parameter they appear.

## 4. EXPERIMENTAL RESULTS ON TETRAEDRON SHAPE

For the tetrahedron punch geometry (first shape in the figure 2), experimental tests are realized, with the same material. Some defaults like plies and wrinkles appear during the forming process. Plies are visible at the bottom of the shape (figure 8), between the blank holders. For this test, the average of the shear angle on one face of the shape is reported figure 9 (a), in function of the punch's stroke. We can conclude, by comparison with the shear behaviour of this material (figure 10) that the locking angle is not reached on this face.

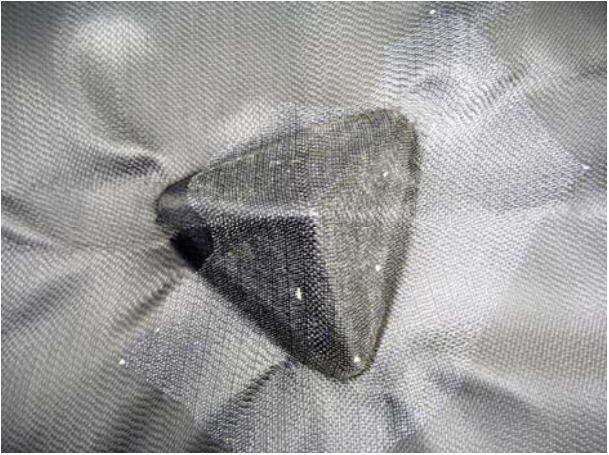

**Figure 8:** Plies on the bottom of the shape

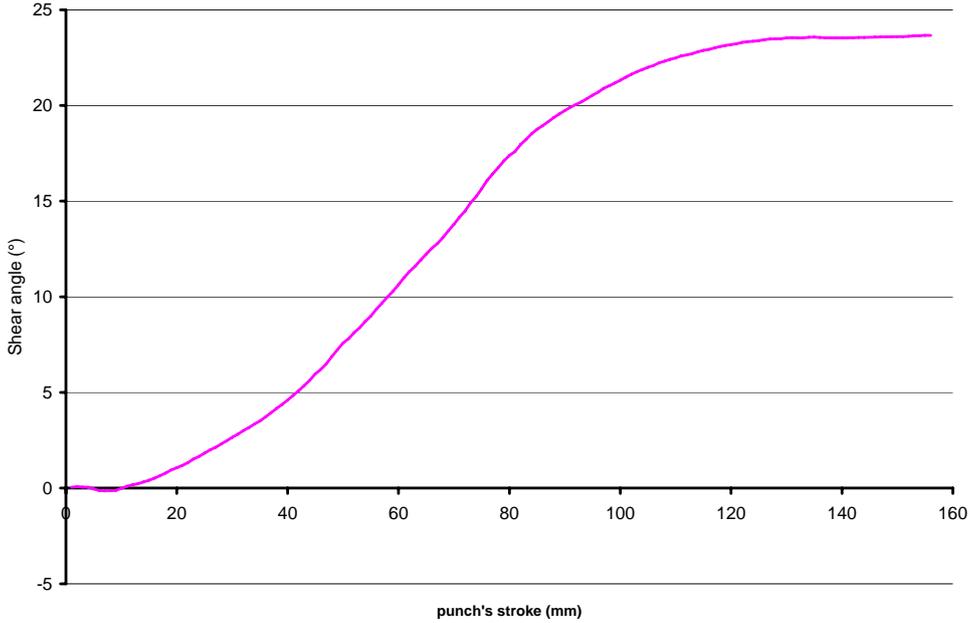

**Figure 9:** Evolution of the shear angle during the process

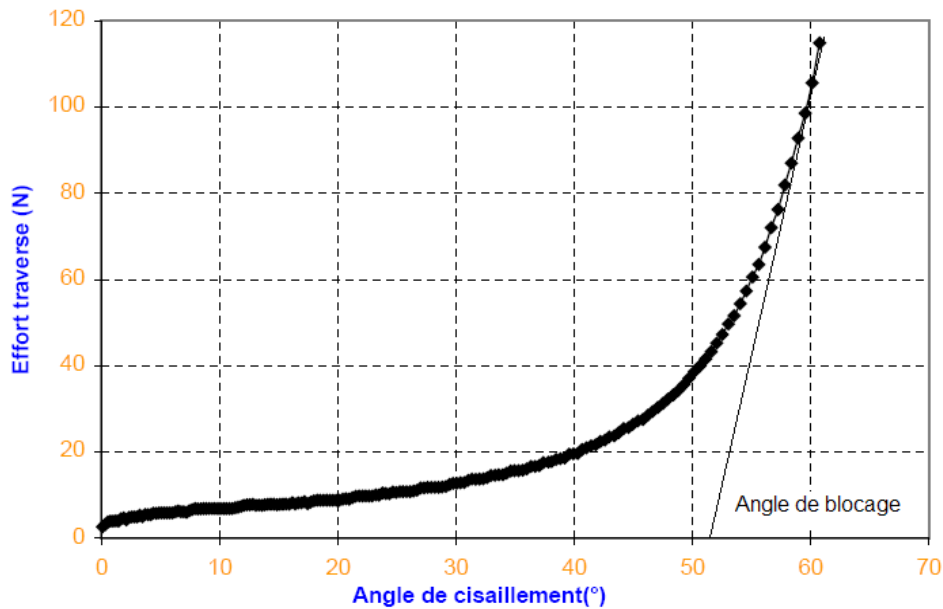

**Figure 10:** Shear Behaviour for the reinforcement

## 5. CONCLUSIONS

This study presented a tool for dry preforms stamping. This tool was developed in collaboration with EADS company. Tests were carried out successfully on different geometry preforms. The post-processing made it possible to obtain the final shape of the preform but also the various intermediary shapes during the test. In addition, this tool also enables us to identify the local behaviour of the fibres.
Future works will be directed towards a basic constitution of experimental data and their correlation with numerical simulations results. In the other hand, parametric studies will be carried out on each preform in order to optimize the process parameters allowing obtaining preforms without defaults.